\documentclass[11pt]{article}
\usepackage{moriond,epsfig}

\def\Journal#1#2#3#4{{#1} {\bf #2}, #3 (#4)}

\def\be{\begin{equation}}
\def\ee{\end{equation}}
\def\bea{\begin{eqnarray}}
\def\eea{\end{eqnarray}}

\begin{document}
\vspace*{4cm}
\title{SYSTEMATIC EFFECTS IN CMB POLARIZATION MEASUREMENTS}

\author{ C. ROSSET }

\address{Physique Corpusculaire et Cosmologie, Coll\`ege de France,\\
11, place Marcelin Berthelot 75231 Paris Cedex 5}

\maketitle\abstracts{The cosmic microwave background polarization is
  rich of cosmological information complementary to those from
  temperature anisotropies. Linear polarization can be decomposed
  uniquely in two components of opposite parities, called $E$ and
  $B$. While $E$ mode allows measurement of cosmological parameters in
  a way independent from temperature, $B$ mode allows to detect the
  primordial gravitational waves produced during inflation, and thus
  to determine its energy scale. However, measuring CMB polarization
  is complicated by foregrounds, whose polarization is poorly known,
  and by systematic effects, which mainly affects $B$ mode measurement
  because of its low level. As an example, we show here the effect of
  beams uncertainty on polarization measurement in the case of the
  Planck HFI instrument, and how we can correct for it.}

\section{Introduction}

Temperature anisotropies have now been detected and measured by many
experiments, most recent results confirming the Gaussianity of
fluctuations, detecting the presence of acoustic peaks in the angular
power spectrum of fluctuations and demonstrating the spatial flatness
of the Universe. This provides compelling evidence that the primordial
perturbations indeed have been generated during an inflationary period
in the very early Universe. The next challenge is now to precisely
measure the polarization anisotropies.

Polarization of cosmic microwave background is produced at the end of
recombination period by Thomson scattering of CMB photons by electrons
of the cosmic fluid. The gradient of fluid speeds induces quadrupole
around electrons, causing emission of linearly polarized light. Linear
polarization can be decomposed into two scalar fields on the sphere,
distinguished by their parity properties: $E$ mode is defined to have
an even parity, while $B$ is odd. Their interest lies in the
difference in physical origins of these two modes: $E$ mode can be
produced by both scalar and tensor modes of primordial density
fluctuations, while $B$ mode can only be produced by tensor
fluctuations. The former, being produced by the same fluctuations as
temperature anisotropies, allows as well the measurement of
cosmological parameters, though it is more sensitive and has different
directions of degeneracy. The concordance of cosmological parameters
obtained from temperature anisotropies and $E$ mode polarization would
be an important test of the cosmological model and, combining both
data, could increase their precision.

On the other hand, the $B$ mode allows the direct detection of the
gravitational waves (or the tensor modes), expected to be produced
during the inflation era. If so, the level of the tensor mode is
linked with the energy scale of inflation, for example in the
slow-roll approximation, by the relation: $E_\mathrm{inflation} =
2\cdot 10^{16} \times \left(\frac{r}{0.1}\right)^{1/4} \mbox{GeV}$,
with $r$ the initial tensor to scalar ratio which can be extracted
from $B$ mode measurement. $B$ mode can also be produced by the
gravitational lensing of CMB photons by large scale distribution of
matter on the way from last scattering surface to us: the polarization
pattern is distorted, so that a fraction of $E$ mode is transformed
into $B$ mode.

The first detection of CMB polarization at one degree angular scale,
at a level compatible with predictions of the standard cosmological
scenario, has been announced by Kovac et al~\cite{kovac2002}, while an
upper limit of 8.4 $\mu$K for the $E$ mode polarization signal at a
sub-degree scale ($l\sim 200$) was established earlier by Hedman et
al~\cite{hedman2002}. More recently, the WMAP team has obtained a
measurement of the temperature-polarization correlation compatible
with expectations on small scales, and bearing on large scale the
signature of unexpectedly early reionization. No significant
constraint on $B$ modes exists yet. The Planck mission, with its full
sky coverage and its polarized detectors in the frequency range
30-353~GHz, will be the first experiment able to constrain
significantly these $B$ modes, and hence to measure them on very large
scales.

The measurement of CMB polarization is complicated by its low level
compared to temperature anisotropies, making it highly sensitive to
both foregrounds, discussed in the next section, and various
systematic effects. As an example, we will expose the problem of the
beam uncertainty for polarization measurement.


\begin{figure}
\centering
\psfig{figure=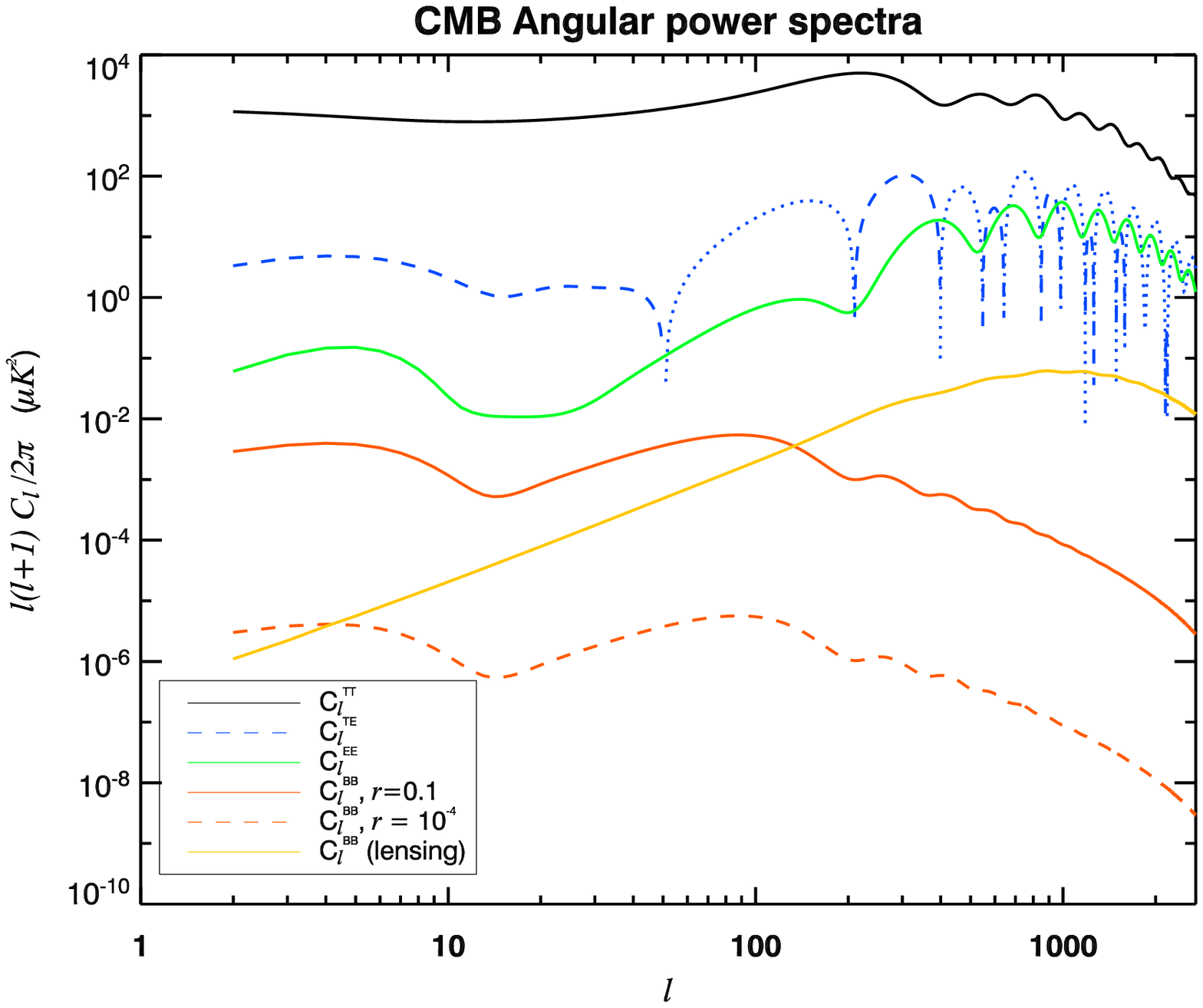,height=1.7in}
\psfig{figure=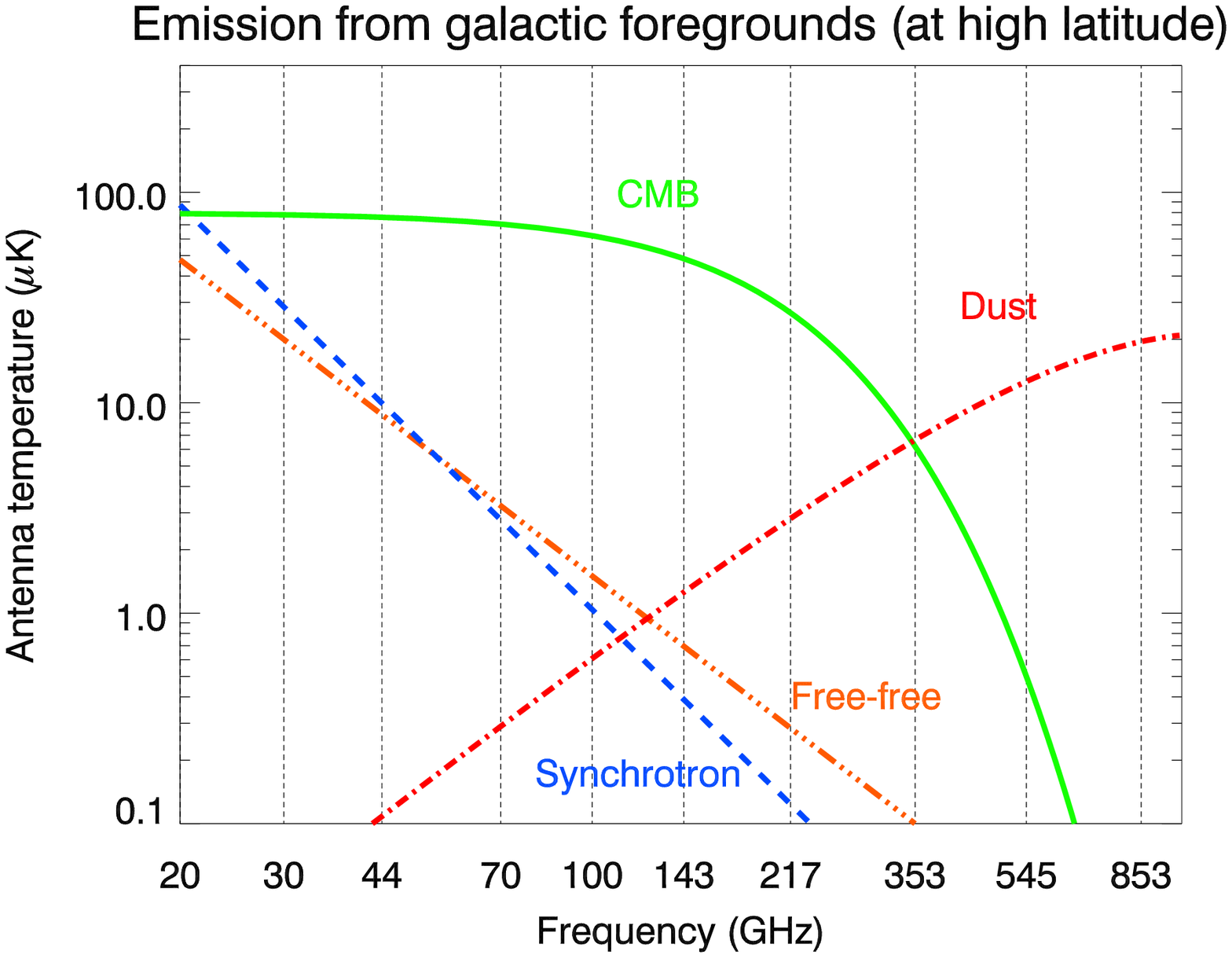,height=1.7in}
\caption{{\it Left:} Angular power spectra of temperature anisotropies
  and $E$ and $B$ mode of polarization. The two contributions from $B$
  mode are separated: tensor mode peaks at $l\sim 90$ while lensing
  contribution peaks at $l\sim 1000$. The tensor mode is shown for two
  different levels of initial tensor to scalar ratio, $r=0.1$ and
  $r=10^{-4}$. {\it Right:} Electromagnetic spectra of intensity of
  the different foregrounds.
\label{fig:pwpolar}}
\end{figure}

\section{Foreground polarization}
\label{sec:forepolar}
The expected level of CMB polarization is of the order of a few $\mu$K
for $E$ mode and around $0.1~\mu$K for $B$ mode. The foregrounds may
very well contaminate maps of polarization. Two main origins of
galactic emission are foreseen around the maximum emission of CMB:
thermal emission of dust at high frequency, which polarization has
recently been measured by Archeops at 353~GHz up to a degree of 20\%
in the galactic plane; and synchrotron radiation at low frequency,
which is expected to be polarized up to 20\%. With a plausible
intensity of a few tenth of $\mu$K, the polarized emission of these
two sources easily dominates the CMB $B$ mode. The distinction between
foregrounds and CMB can be done through the difference in
electromagnetic spectrum of the sources.

\section{Systematic effects from beams}
\label{sec:systeff}
The weakness of the signal to measure makes the $B$ mode particularly
sensitive to various systematic effects. Beside usual systematic
effects occurring in CMB experiments, such as $1/f$ noise, a whole
class of them is specific to polarization, as its measurement involves
the differences of signals. Indeed, linear polarization is
characterized by two Stokes parameters, $Q$ and $U$, defined as the
differences of intensity through two polarizers at 90 degrees one from
the other ($0^\circ$ and $90^\circ$ for $Q$, and $45^\circ$ and
$135^\circ$ for $U$). Any differences between detectors combined to
measure $Q$ and $U$ may result in a spurious polarization measurement,
usually by transforming $E$ mode into $B$ mode or temperature
anisotropies into both $E$ and $B$ polarization, as $T \gg E \gg B$.

The Planck High Frequency Instrument will measure polarization using
Polarization Sensitive Bolometers (PSB) associated by pair inside one
horn, each measuring the intensity for one direction of
polarization. The difference of the signal from two PSB inside one
horn thus gives a combination of $Q$ or $U$ in the frame of the focal
plane. In order to get both $Q$ and $U$ in the sky reference frame, we
need to combine the measurements from two horns.

\begin{figure}
{\centering\psfig{figure=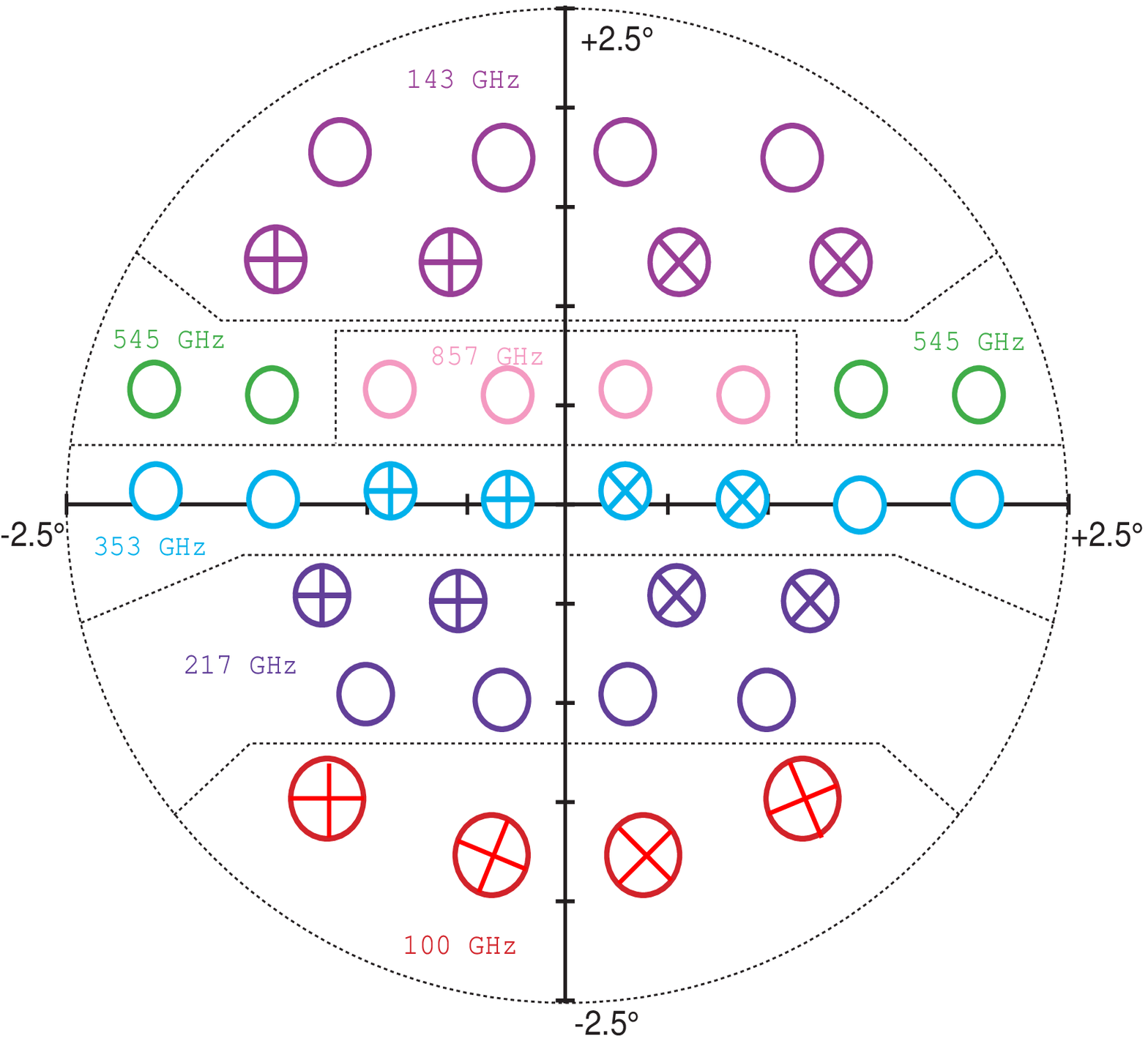,height=1.45in}
\psfig{figure=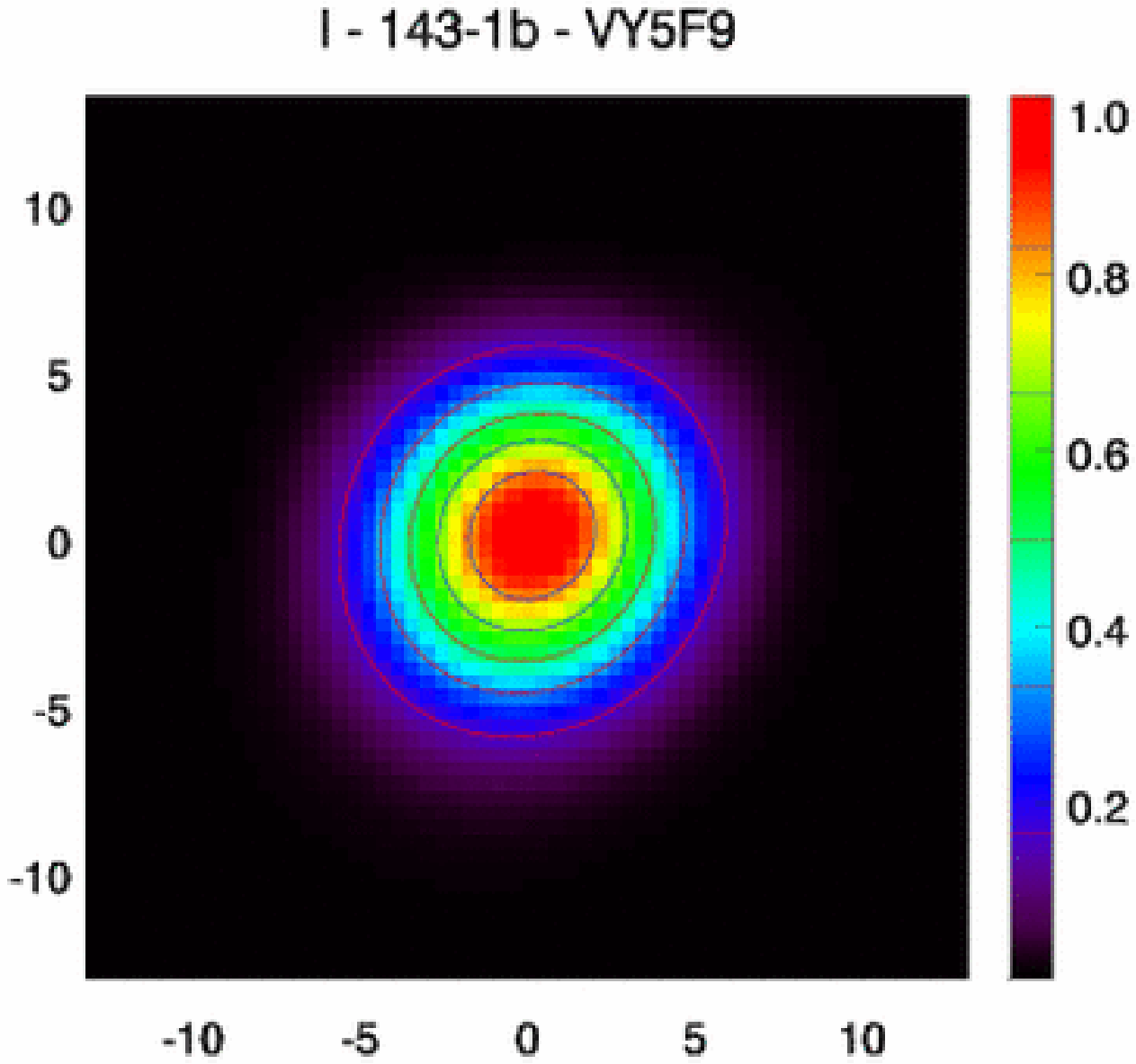,height=1.55in}
\psfig{figure=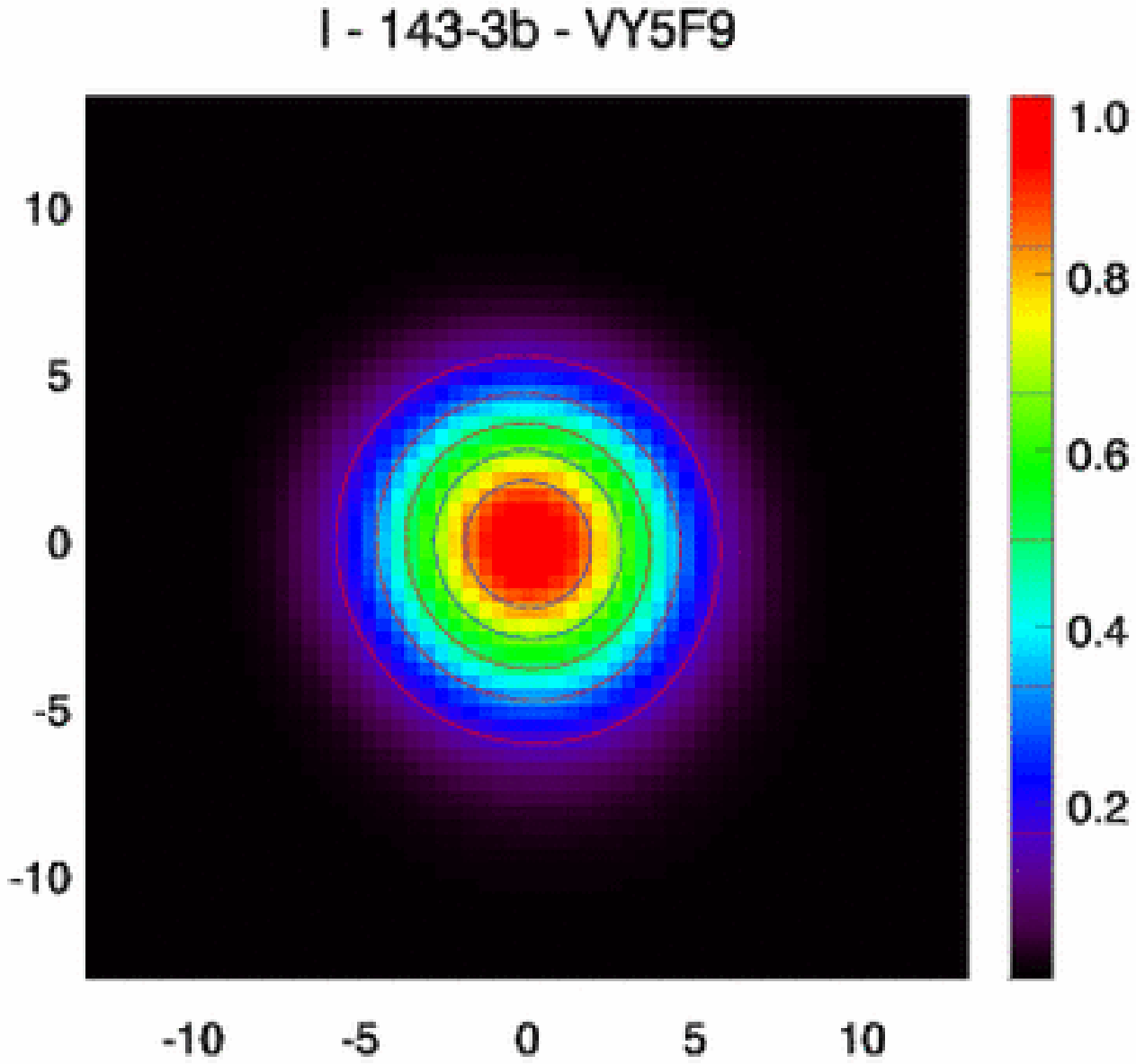,height=1.55in}
\psfig{figure=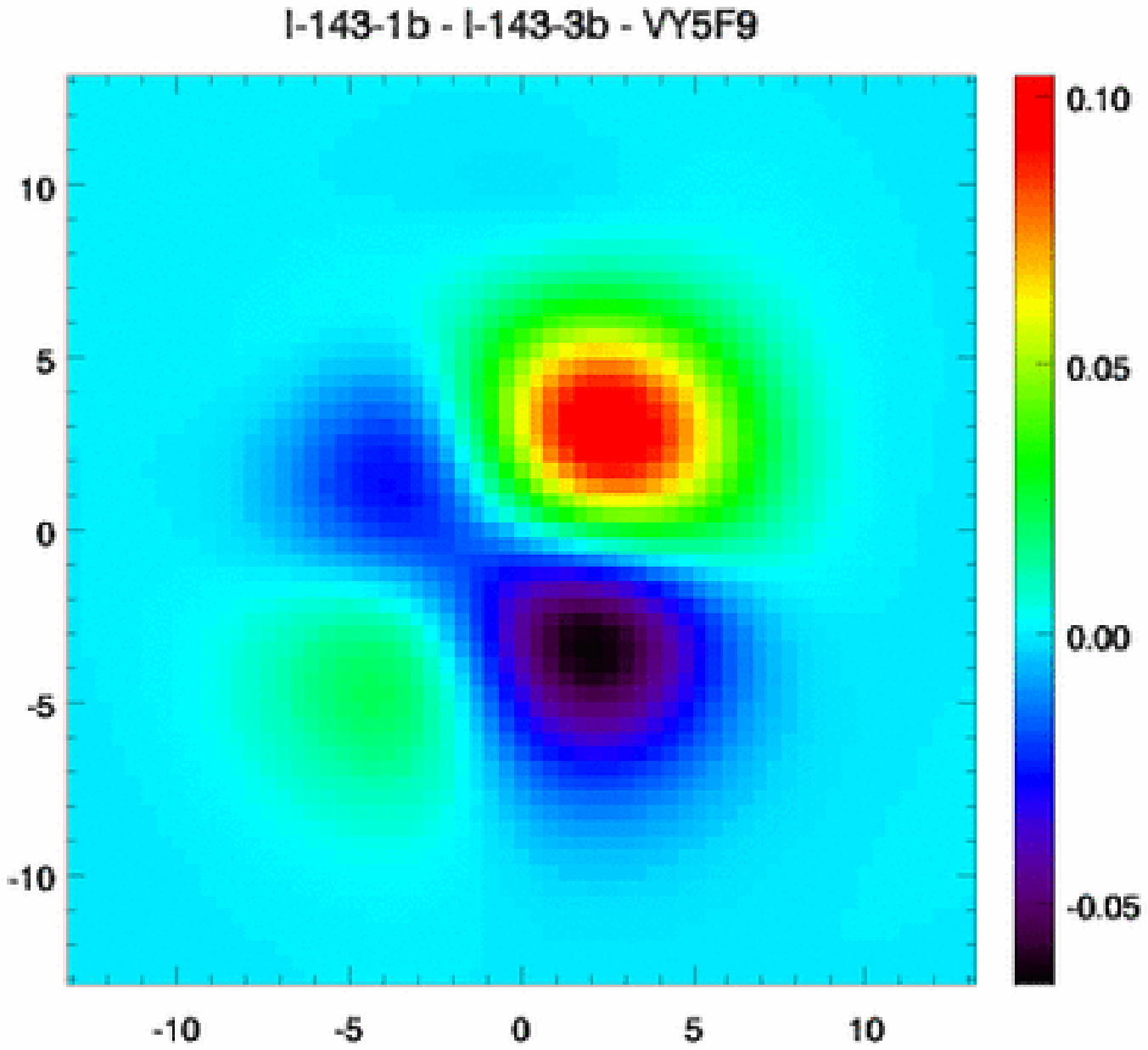,height=1.45in}}
\caption{{\it Left:} Focal plane of the Planck High Frequency
  Instrument. {\it Center:} the beams computed for two horns,
  elongated in different directions. {\it Right:} the difference
  between the two beams of two different horns, which is up to 10\% of
  the beam peak.
\label{fig:beams}}
\end{figure}

An electromagnetic simulation of the optical system of Planck,
including the telescope and the horns, done by
V. Yurchenko~\cite{yurchenko}, shows that the beams of the horns are
elliptical (see Fig.~\ref{fig:beams}). The difference between beams of
different horns is then up to 10\% of the beam peak, while the
difference of intensity beams of the two detectors within the same
horn is less than 0.5\%.

\begin{figure}[h]
\centering\psfig{figure=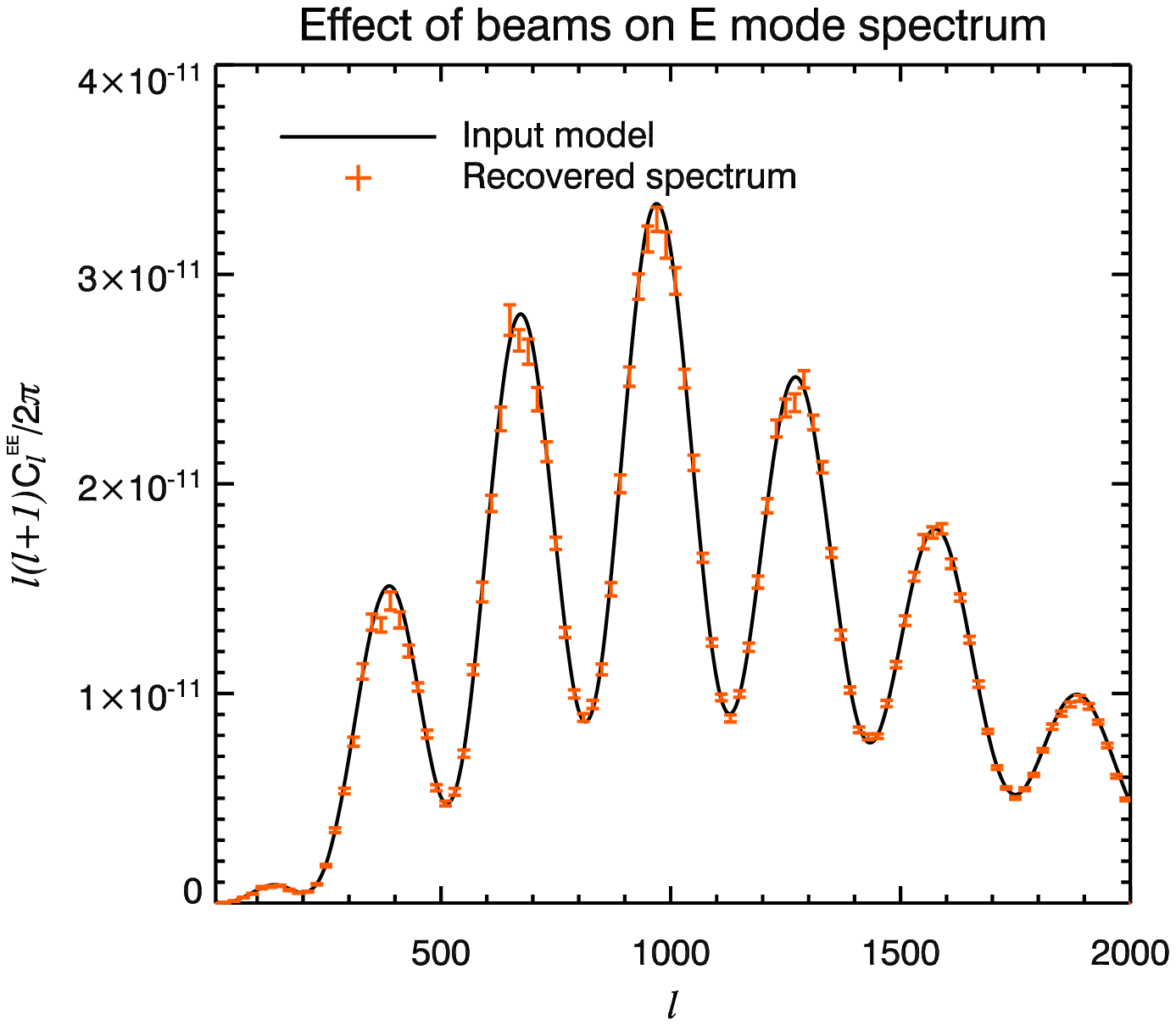,height=1.8in}
\psfig{figure=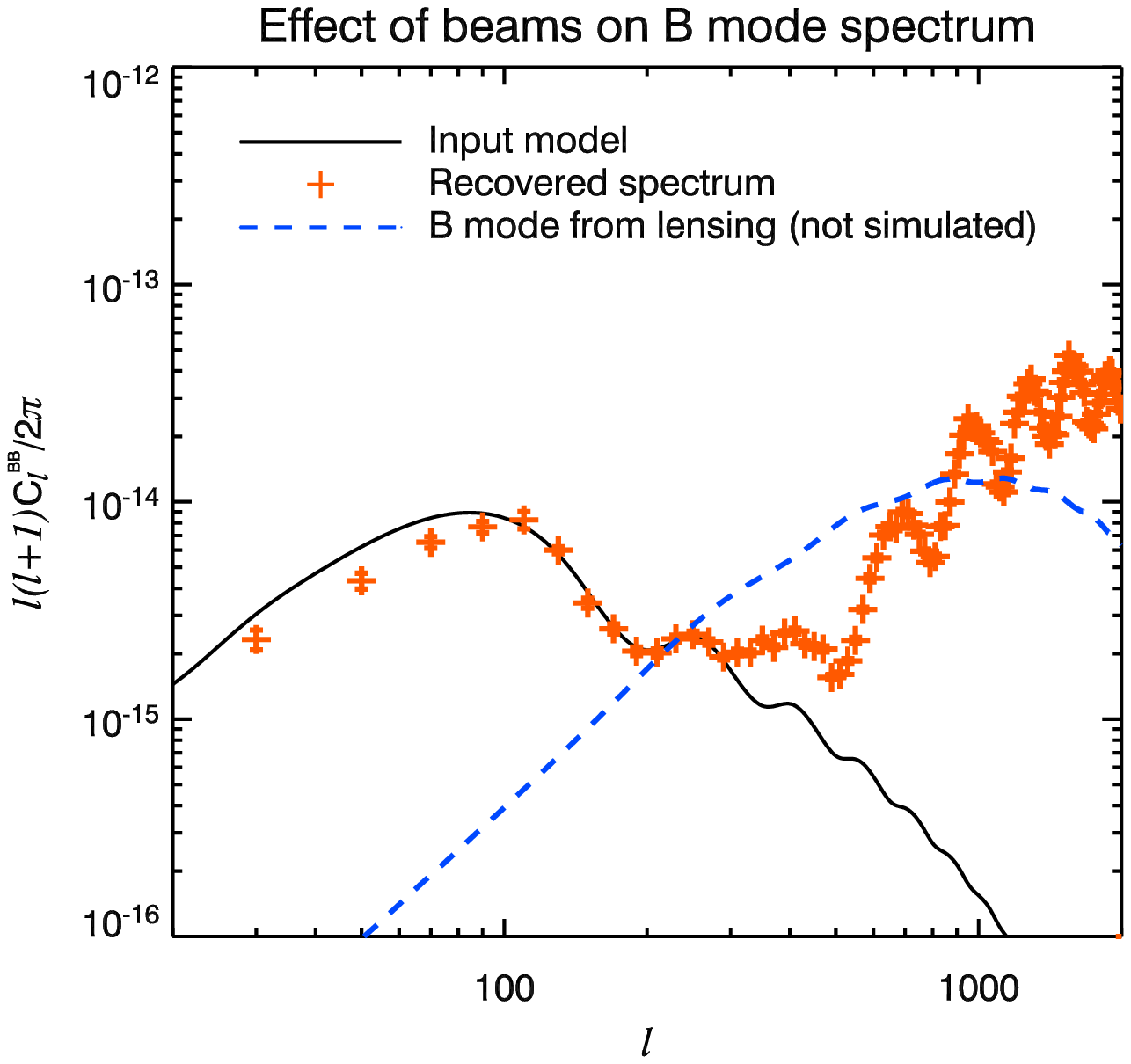,height=1.8in}
\caption{Recovered polarization power spectra with realistic simulated
  beams (see text and figure~\ref{fig:beams}). $E$ mode is recovered
  with a precision 0.1\%, while $B$ mode is affected at low angular
  scales by a temperature and $E$ mode leakage due to beam
  differences.
\label{fig:recpw}}
\end{figure}

We have estimated the effect of such beams on the measurement of CMB
polarization power spectra by Monte-Carlo, using these simulated
beams, on plane maps with a scanning strategy realistic for
Planck. The power spectra of temperature and $E$ mode are recovered
with an error less than 0.1\%, while the $B$ mode is biased by a
spurious polarization mainly coming from a $E$ mode leakage, but also
from a temperature leakage. The spurious $B$ signal overcomes the CMB
signal from $l\sim 300$ and the lensing $B$ mode from $l\sim 700$ (see
Fig.~\ref{fig:recpw}).

\begin{figure}[h]
\centering\psfig{figure=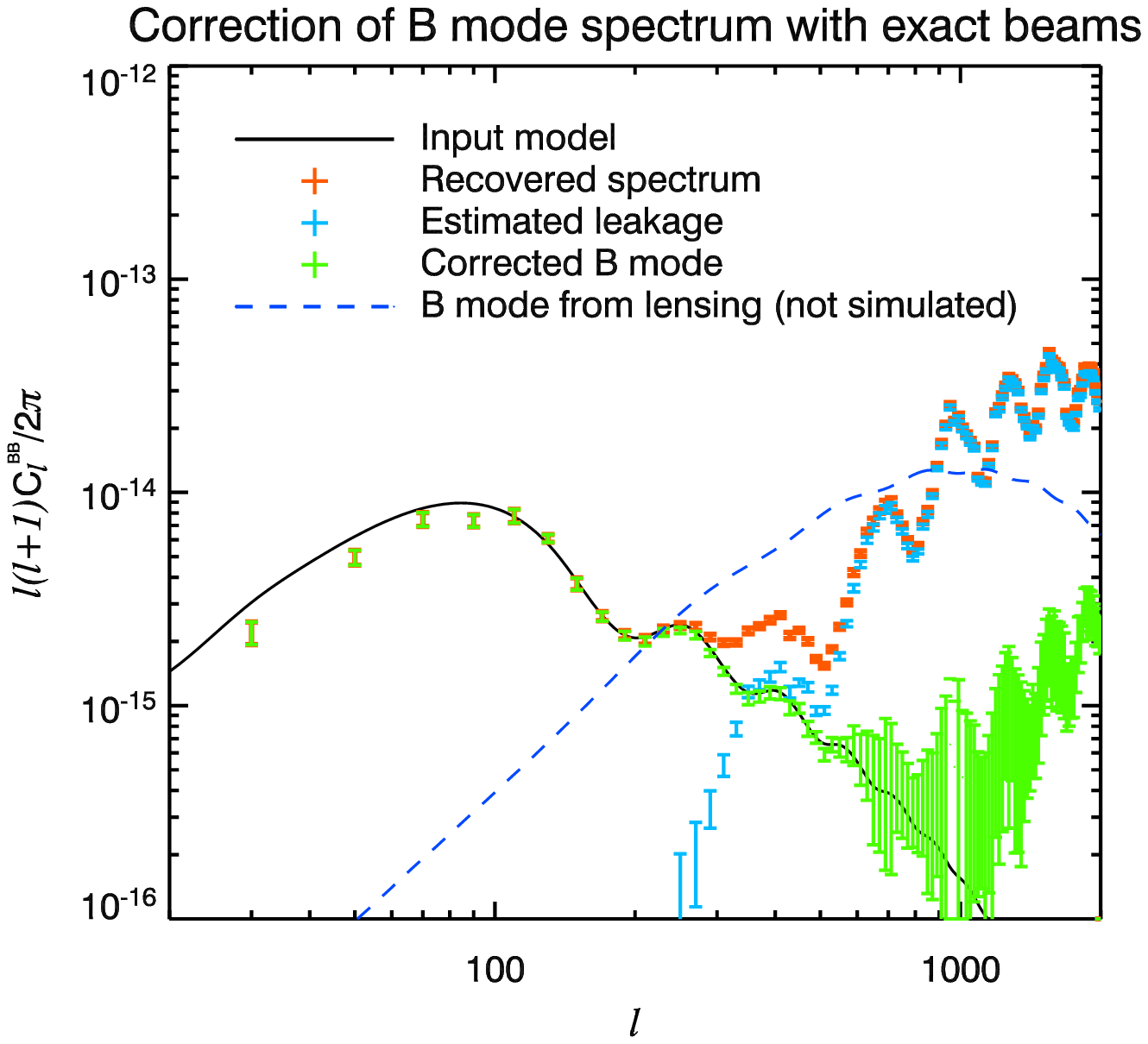,height=1.8in}
\psfig{figure=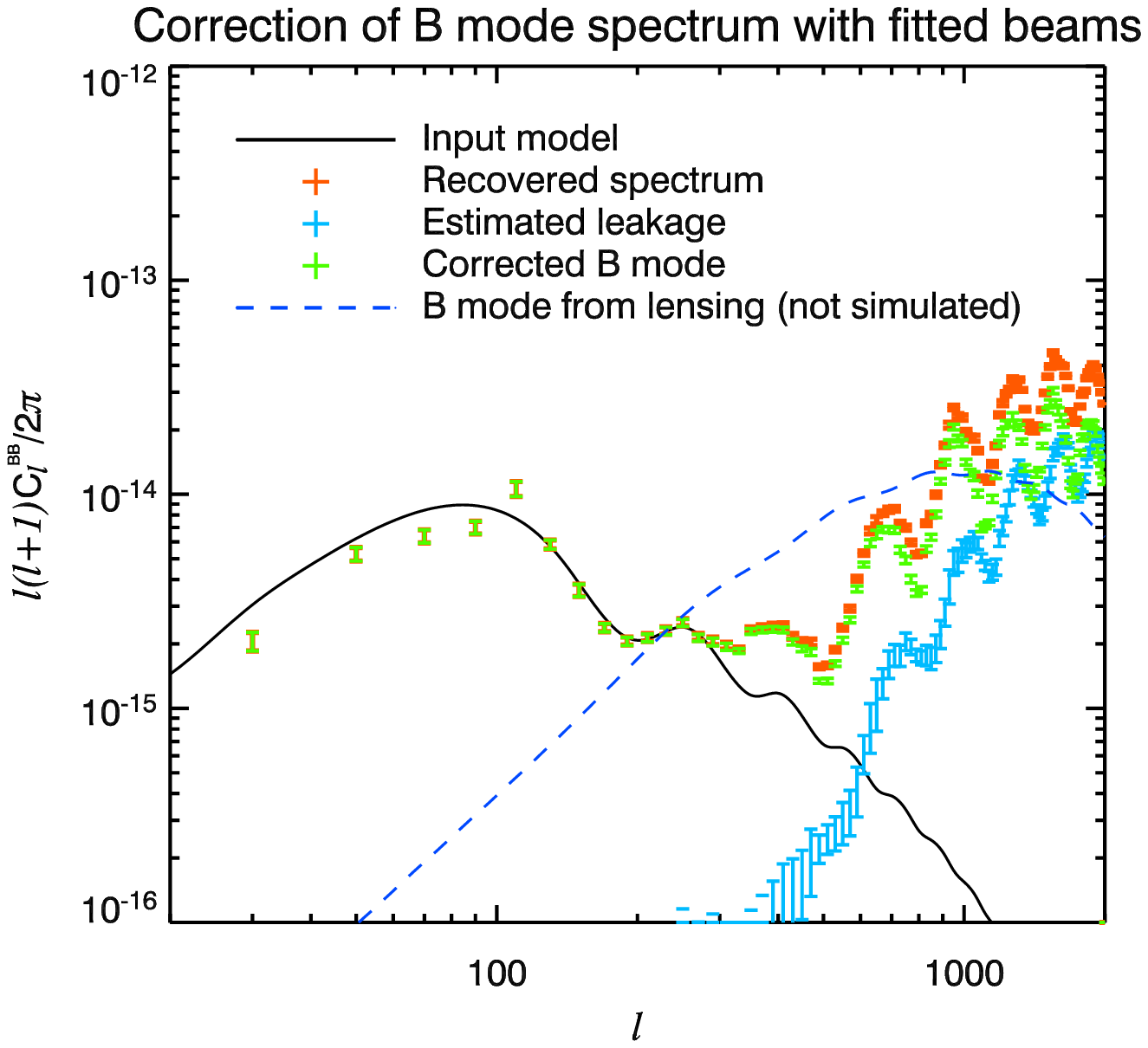,height=1.8in}
\caption{Correction of the $B$ mode using exact beams ({\it left}) or
  Gaussian fit of the beams ({\it right}) with the method described in
  the text.
\label{fig:corrpw}}
\end{figure}

However, it is possible correct for this effect by estimating the bias
induced by the beams difference if we have a precise enough knowledge
of them: we can use the recovered temperature and $E$ mode maps as
input sky and simulate the instrument using the {\it known} beams. The
output of this simulation contains some $B$ mode, not present
initially, which is an estimation of the observed spurious $B$ mode
(see Fig.~\ref{fig:corrpw}). The efficiency of this correction
strongly depends on how well the beams are known. As an example, if we
approximate the beams by asymmetric two dimensional Gaussians, which
are up to 2\% different from the exact beams, the estimation of the
spurious $B$ mode is too low, making the correction inefficient.

\section{Conclusion}
\label{sec:conclusion}
Measuring the faint CMB polarization signal is challenging because of
the sensibility to various contaminations: astrophysical, as the
foregrounds may overcome the CMB signal, and instrumental, because of
the differential nature of the polarization. We have shown here the
effect on polarization power spectra measurement due to differences in
the beams, using realistic simulations for Planck. Other systematics
are possible and important for polarization measurement with Planck,
such as the relative calibrations or the time constants of
bolometers. Specific methods have to be developped to take these
effects into account and correct for them, particularly for the $B$
mode power spectrum reconstruction.



\section*{References}
\begin{thebibliography}{99}

\bibitem{kovac2002} J. Kovac, E. M. Leitch, C. Pryke, J. E. Carlstrom,
  N. W. Halverson \& W. L. Holzapfel \Journal{{\em
  Nature}}{420}{772}{2002}

\bibitem{hedman2002} M. N. Hedman, D. Barkats, J. O. Gundersen,
  J. J. McMahon, S. T. Staggs, \& B. Winstein \Journal{\em
  ApJ}{573}{L73}{2002} 

\bibitem{yurchenko} V. Yurchenko, J. Murphy \& J. M. Lamarre in {\em
  Proceedings of 3rd ESA Workshop on Millimeter Wave Technology and
  Applications} ed. J. Mallat et al pp. 187-192 (2003)

\end{thebibliography}
\end{document}